# Enhanced Spin-Orbit Interaction of Light through Complex Fresnel Reflection

**Dileep Kumar Upadhyay and Nirmal K. Viswanathan**

School of Physics, University of Hyderabad, Gachibowli, Hyderabad 500046, Telangana, India

nirmalsp@uohyd.ac.in

**Abstract:** *Spin–orbit interaction* (SOI) of light manifests as the generation of a spin-dependent vortex beam and the spin-Hall effect when a spin-polarized beam interacts with an optical interface at normal and oblique incidence, respectively. The weak SOI limits its widespread use for quantifying physical quantities in various applications. We demonstrate a significantly enhanced SOI effect, approaching the strong-coupling regime, by retroreflecting a nonparaxial focused beam of light at a multilayer dielectric mirror. The large difference in angle-of-incidence-dependent Fresnel reflection coefficients $(r_{TE}, r_{TM})$ introduced by the mirror simultaneously enhances the spin-orbit-converted vortex-beam amplitude and the spin-Hall effect in the reflected beam. Near-field superposition of the orthogonal polarization components in the reflected beam results in a spatially nonuniform beam field with $C$, $V$, and $L$ polarization singularities. These topological features are mapped and are used to demonstrate the transverse and 3D spin-Hall effect in the focal region. The superposition of enhanced optical near-fields and the lossless broadband nature of the mirror reflectivity are expected to enable a variety of polarization topologies and to tailor their features via designed mirror reflectivities.

**Introduction:**

Spin-orbit interaction (SOI) of light is a generic mechanism by which the spin angular momentum (SAM) and orbital angular momentum (OAM) degrees of freedom, or alternatively the intrinsic angular momentum (IAM) and extrinsic angular momentum (EAM), are nonseparably coupled [1 – 3]. The resulting non-uniform state of polarization has led to the observation of a variety of fundamental effects, including spin-to-orbital AM conversion (SOC), spin-Hall effect (SHE), orbital-Hall effect (OHE) [4 – 8], their reciprocals [9, 10], as well as higher-order effects [11 – 15]. The weak coupling mechanisms in typical macroscopic optical systems led to the first proposal of the spin-Hall effect and its experimental demonstration [16, 17], using the quantum weak measurement (WM) method [18, 19]. This has since continued to dominate the measurement and quantification of ultrasmall phase-polarization variations arising from the SOI of light [6, 19]. On the other hand, the AM components become strongly coupled due to enhanced anisotropy and inhomogeneity in the optical near-field arising from strong spatial confinement and subwavelength-scale interactions with the non-propagating evanescent field [1, 20 – 24]. The strong SOI effects are typically measured using near-field optical probes, such as an optical fiber tip of a near-field scanning optical microscope (NSOM), or by scanning a nano scatterer to reconstruct the optical near field [25 – 32].

The question that arises in this context is how to transport the strongly coupled optical near-field to the propagating optical far-field to enable applications. This is achieved by coupling the SAM of the incident light via a sub-wavelength scatterer to the evanescent wave of a surface plasmon polariton or waveguide modes [33 – 35]. The SAM-controlled unidirectional coupling to directional surface plasmon polaritons [33, 35], or guided-mode propagation via evanescent-wave coupling of nanoparticle-scattered radiation with near-100% directionality [34], provides exciting approaches towards classical and quantum spin-photonic devices [36, 37]. Both these demonstrations are enabled by a significant increase in SOI strength due to the strong transverse confinement of optical fields. Enhanced SOI enabled by nanoparticle scattering [24, 25, 27], scattering in planar photonic crystal microcavities [38, 39], and scattering at nanostructured metasurfaces [28 – 32, 40, 41] has since spurred an explosion of research activity. Additionally, coupling of light's AM in exotic optical structures such as three-dimensional microcavities [42, 43], in moving matter due to the light-dragging effect [44], and in specialty optical fibers [45, 46] has also been shown to enhance the SOI. Besides, light-matter interaction-related enhancement in the SOI has also been explored [47 – 51]. All these have led to the observation of optical vortex structures arising from SOC and the photonic SHE, stemming from the underlying geometric phase and the conservation of AM [1, 52 – 54].

Due to weak SOI, the fundamental SOC and SHE were proposed and demonstrated in various optical systems, and various mechanisms were used to interpret the results [1, 4 – 7, 55 – 60]. Only recently have these two effects been unified within a single framework [61 – 63]. An angle-of-incidence-dependent topological phase transition from vortex generation to the spin-Hall effect was revealed using a designed metasurface [61]. The primary objective behind the metasurface design was to enhance the generation efficiency of the abnormal (vortex) mode by tuning the anisotropic response along the in-plane ($xy$) and the propagation ($z$) directions. This enhances the difference in the optical responses of the TE and TM polarization components, thereby converting a greater fraction of the input beam's normal mode into an abnormal mode. Further, the optical vortex generated by focusing a circularly polarized Gaussian light beam has the same SOI origin as passing a light beam through an optically thin dielectric slab, with the momentum-dependent Pancharatnam-Berry phase as the mechanism [55].

One largely overlooked but significant aspect of the SOI is the emergence of polarization singularities, which are topological defects in spatially inhomogeneous optical fields. Polarisation singularities (PS), discovered by Nye and Hajnal [64, 65] and extensively researched and written about in review articles and books [66 – 76], are the fundamental electromagnetic structures of light. In a generic spatially inhomogeneous light field with spatially varying polarization ellipticity and ellipse orientation, the most basic structures are the C-singularities, where the polarization is purely circular, and the L-singularities, where the polarization is purely linear [65]. The C-point PS corresponds to equal orthogonal field amplitudes with $\pi/2$ phase difference between them. In the transverse cross-section of the measurement plane, this corresponds to $(E_x \pm iE_y)$. For a generic 3D field, the two field components can be any pair among $(E_x, E_y, E_z)$ which, for radially symmetric systems, such as in symmetric focusing, become transverse and longitudinal field components $(E_\perp, E_\parallel)$. Identification of C-points in a plane involves measurement of Stokes parameters, with the condition $s_3 = \pm 1$ imposed on the normalized third Stokes parameter. L-line singularities are places where the polarization is linear, and $s_3 = 0$, where right-elliptical polarization transitions to left-elliptical polarization.

Considered as organizing structures, the state of polarization of light around these singularities organizes into various topological structures. Polarization filtering of these fields reveals the underlying geometric-phase structure, including point and line phase singularities, associated Poynting-vector structures, and energy-flow patterns [77 – 80]. In the context of the SOI of light, the presence of a C-point singularity indicates spin-dependent vortex structure around a phase singularity, and the L-line singularity indicates the spin-Hall effect of light, respectively, due to SAM-to-intrinsic OAM and spin-to-extrinsic OAM conversion. While these effects are dominant in the optical near field due to the strong coupling of light’s AM, they were largely ignored in the far field, as their contribution to beam-field modifications is very weak.

A spatially confined optical beam field provides an ideal setting for exploring a variety of nontrivial phase–polarization effects arising from enhanced SOI of light [1, 81 – 86]. For optically thin-slab systems, it is proposed that the SOC efficiency can be enhanced by either increasing the beam's nonparaxiality or increasing the difference between the TE and TM polarization components. Is it then possible to have an equivalent system operated in reflection mode, where the Fresnel reflection coefficients $(r_{TE}, r_{TM})$ are sufficiently different to significantly enhance the SOI effects, using which we can demonstrate both the SOC and photonic SHE in the same configuration, and also show that these effects happen respectively around the C-point and L-line singularities. A high-reflectivity multilayer dielectric mirror is one such system [87, 88], using which we demonstrate enhanced SOC from a circularly polarized input Gaussian beam, and the photonic SHE via Stokes polarimetry measurements. The reflected normal and abnormal mode amplitudes are sufficiently high that on-axis superposition of these in the near-field results in the appearance of C- and L- type PS in the beam cross-section for circular and linear input polarization. Using the C-point PS as a subwavelength spatial marker, we track the polarization variation in the focal region by scanning the mirror along the propagation direction. We provide a clear demonstration of the SHE in 3D [81] where the $s_3$ Stokes component flips sign due to the Gouy phase, while the AM content of the beam remains the same across the focal plane.

Here, we demonstrate a method to convert a uniformly polarized Gaussian input beam into a topologically structured output beam with PS by retroreflecting the focused beam off a dielectric multilayer mirror positioned in the focal region. The superposition of all three nonzero field components in the nonparaxial focal region, upon interacting with the 3D multilayer structure, results in strong SOI, which generates a structured light beam whose characteristics vary dynamically as the mirror is translated across the focal region. Using this system, we demonstrate both the fundamental SOI effects, the SOC and photonic SHE, respectively, around C-point and L-line singularities, for circular- and linear-polarized input beams. By translating the mirror across the focal plane, we also observe changes in their behavior in 3D. The anisotropy and inhomogeneity arising from the complex Fresnel reflection coefficients of the multilayer dielectric mirror strengthen the near-field coupling, enabling Stokes polarimetry measurements to characterize various phase-polarization effects in the far field. Our method thus provides a simple and novel approach to generating PS beams and manipulating their characteristics via spatially varying birefringence, owing to the mirror's longitudinal periodicity. Compared with a single nano-scatterer used to map nonparaxial fields or a designed array of nano-scatterers on a metasurface to manipulate the transverse beam profile, our method leverages the longitudinal degree of freedom to both study and manipulate optical near-field effects, thereby converting them into structured paraxial beams of light [89].

**Theoretical details:**

Several SOI effects arising from enhanced phase-polarization modifications in the retroreflected, nonparaxial, focused beam are simulated and confirmed by experimental measurements. The theoretical simulations and the experimental measurements presented here are carried out for circular and linear-polarized Gaussian beams from a laser of wavelength $\lambda$=632.8 nm, focused by a microscope objective lens with a numerical aperture (NA) of 0.95. The focused beam is retroreflected by a multilayer high-reflectivity dielectric mirror kept at the focal plane, perpendicular to $\langle \boldsymbol{k}_0 \rangle$, the mean propagation direction. The retroreflected beam passes through the same high-NA lens, is collimated, and is imaged onto a CCD camera, enabling spatially resolved measurements of the state of polarization and weak polarimetry.

Theoretical simulation of the nonparaxial focused beam is carried out following the Richards–Wolf formalism [90]. Upon interaction of the focal beam-field at the mirror, the enhanced retroreflected optical beam-field is given by

Equations (2)

In the above equations, $f$ is the focal length of the high-NA lens, $r_s$ and $r_p$ are the Fresnel reflection coefficients of the mirror, $f'$ imaging lens focal length, $f(w)$ pupil function to ensure beam apodization $z$, the propagation distance, $z_0$ defocused length and $J_{0r}$ and $J_{2r}$ are the Bessel functions. Using the complex transverse optical field components $(E_x,\ E_y)$ or the corresponding intensities $I_i(\theta,\phi); i = 1-6$, with $(\theta,\phi)$ quarter-wave plate angle and analyzer angle, respectively, the spatially non-uniform state of polarization (SoP) of the beam is calculated using the four Stokes parameters $(S_i); i = 0-3$:

$S_0 = |E_x|^2 + |E_y|^2 = I_1(0^o,0^o) + I_2(90^o,0^o)$; $S_1 = |E_x|^2 - |E_y|^2 = I_1(0^o,0^o) - I_2(90^o,0^o)$; $S_2 = E_x E_y^* + E_x^* E_y = I_3(45^o,0^o) - I_4(135^o,0^o)$; $S_3 = -i(E_x^* E_y - E_x E_y^*) = I_5(45^o,90^o) - I_6(135^o,90^o)$ **(3)**

We consider only the pixel-wise polarized component in the output beam by normalizing the Stokes parameters using $s_i = \frac{S_i}{pS_0}; i = 1,2,3$, where, $p = (\sqrt{S_1^2 + S_2^2 + S_3^2})/S_0$ is the degree of polarization (DoP). In the spatially nonuniform elliptically polarized beam cross-section, we identify C-point singularities satisfying conditions $s_1 = s_2 = 0$; $s_3 = 1$, being points of circular polarization with undefined polarization azimuth and L-line singularities where $s_3 = 0$ which are locations with undefined handedness. The C-points are characterized by an index $I_c = \frac{1}{2\pi}\oint d\alpha$, where $\alpha$ is the orientation angle of the major axis of the polarization ellipse. For the integral carried out on a loop drawn around the C-point, $I_c$ takes values of $\pm 1/2, \pm 3/2$ .... From this point of view, it is convenient to use the complex Stokes field $s_{12} = s_1 + is_2$, which along with $s_3$ are used to locate and characterize the polarization singularities. The C-points are located at the zeros of the $s_{12}$ field, and the L-lines are determined by locating the zeros of $s_3$. The orientation of the polarization ellipse at each point is determined using $2\Psi = arg(s_{12})$. A positive value of $s_3$ indicates right-handedness, while a negative value indicates left-handedness. The orientation of the major axes of the polarization ellipses is then connected via streamlines to identify polarization topologies. The streamlines in the ellipse orientation field given by $\Psi = \frac{1}{2} tan^{-1}\left(\frac{s_2}{s_1}\right)$ are the family of solutions of the differential equation $dy/dx = sin\chi/cos\chi$, which forms the topological structure around the singular point. The abrupt turn of the streamline at the C-point singularity implies an indeterminate orientation of the polarization ellipse. The streamline pattern around the C-point is used to identify and visualize the *lemon*, *star*, or *monstar* pattern of index $\pm 1/2$.

The enhanced SOI effects from the proposed combination of nonparaxial focusing and multilayer mirror reflection enable an in-depth investigation of a range of phenomena using a single system. This includes an investigation of spin-to-orbital AM conversion, the spin-Hall effect, the orbital-Hall effect (not demonstrated here), and transverse spin dynamics (recently demonstrated by us). Moving the mirror across the focal region allows us to study these effects along the propagation direction and carry out 3D mapping of them. We calculate and plot the variation in phase, polarization ellipticity, and ellipse orientation across the beam cross-section as a function of the input beam polarization and mirror position along the propagation direction. The overall polarization variation in the beam cross-section, as a function of input beam polarization and mirror position in the focal region, allows us to identify C-point and L-line polarization singularities and the *lemon, star, and monstar* polarization topologies around them. These are used to quantify variations in angular momentum and energy flow.

**Experimental details:**

The schematic of the experimental setup is shown in Fig. 1. Gaussian beam from the laser source (He-Ne, $\lambda = 632.8$ nm) passes through a Glan-Thompson polarizer (P1) and a zero-order half-wave plate (HWP), ensuring a vertically polarized input beam. The vertically polarized beam is reflected by the 50/50 beamsplitter (BS), passes through the objective lens (infinity-corrected Olympus Plan-C N, 100X/1.25 oil), and is tightly focused. The diameter of the incident beam is approximately 1.6 mm, and the diameter of the lens aperture is 3mm, resulting in a fill factor of 53%. The tightly focused beam is retroreflected by the multilayer dielectric mirror (BB1-E02, Thorlabs, USA) kept at the focal plane. The mirror mounted on a nano-positioner stage (Thorlabs, USA) can be translated within the focal region to map phase-polarization changes. The retroreflected beam from the mirror (kept at the focal plane) passes through the same objective lens, gets collimated, and is transmitted through the beamsplitter (BS). Nonparaxial focusing, followed by a complex Fresnel reflection at the mirror, significantly enhances the polarization inhomogeneity of the retroreflected beam by superimposing the normal and anomalous polarization components. As compared to the 3 – 4 orders of magnitude different orthogonal field components $\left(E_x,\ E_y\right)$ or $(E_{\sigma^+},\ E_{\sigma^-})$ at the nonparaxial focus, the mirror-reflected orthogonal field components are only an order of magnitude different. The resulting output beam thus exhibits enhanced phase-polarization inhomogeneity due to the superposition of the orthogonal field components. This enables measurement of changes in the beam cross-section as the mirror is translated in the focal region using spatially resolved Stokes parameter measurements, allowing us to experimentally map all SOI-related effects in 3D. The Stokes parameters of the output beam are measured by passing it through different fixed orientations of the quarter-waveplate (QWP) and polarizer (P2) combination and measuring the intensity of the beam captured with a CCD camera (Kiralux 2.3 MP Color CMOS Camera, USB 3.0 Interface).

**Results and Discussion:**

It is well known that at the focal plane, the intensity of the Gaussian beam is a maximum and its waist size is a minimum. However, the polarization characteristics of the Gaussian beam in the nonparaxial focal region have been investigated to a limited extent, either using a near-field scanning optical microscope (NSOM) [28] or by scanning a nanoparticle [27]. Especially important to us here is the phase-polarization variations that the beam undergoes in the focal region. For a nonparaxially focused linearly polarized Gaussian input beam, the spin component of polarization, $S_3$ = 0, at the focal plane. However, because our measurement of polarization changes is performed after complex Fresnel reflection at the mirror, we observe a small $S_3$ component at the focal plane, whose magnitude varies with the mirror's position. To understand this, Stokes parameter measurements are taken for each mirror position, shown in Fig. 2 (inset (a) – (c)) for 3 mirror positions. The characteristic four-lobe pattern with alternating $S_3$ corresponds to azimuthal spin separation, arising due to SOI of light [89]. The experimentally measured $S_3$ are simulated, and the results are shown in Fig. 2 (inset (a′) – (c′)).

From the spatially resolved $S_3$ we obtain the spatially-averaged magnitude $\langle|S_3|\rangle$, which is plotted as a function of the mirror position in Fig. 2 (d). The position of the mirror at which $S_3$ is a minimum is designated as the focal plane in our measurements, and all measurements of changes in the beam characteristics before and after focus are referenced to this mirror position. The minimal incremental movement i.e.0.8 $\mu m$ and repeatable position accuracy i.e. 15 $\mu m$ of the translation stage limits the longitudinal resolution of our measurements. Nevertheless, we observe significant changes in the $S_3$ value due to the mirror's enhanced reflectivity. It is important to note here that $S_3$ variation is asymmetric with reference to the focal plane ($z = 0\ mm$). The difference between the two $S_3(z)$ curves is attributed to the mirror model used in simulations as compared to the one used in the experiments. An approximate multilayer mirror model, with complex reflection characteristics as shown in Fig. 1, is used due to the unavailability of exact information on the multilayer coatings used in the mirror.

Next, with the mirror kept fixed at the $z = 0\ mm$ position, a more detailed phase-polarization analysis is performed on the measured Stokes parameters. For the linearly polarized input beam, the Stokes images for the $(I_5,\ I_6)$ intensity measurements shown in Fig. 3 (a), (b) are used to calculate the third Stokes parameter using $S_3 = I_5(45^o, 90^o) - I_6(135^o, 90^o)$ (Fig. 3 (c)). The yellow and blue regions, respectively, correspond to right and left elliptical polarization, and they alternate azimuthally, and the circle where these two join corresponds to linear polarization, where $s_3 = 0$ which are locations with undefined handedness and hence L-line singularity. The spatial positions where the ellipse orientation is undefined, and corresponding to $(\pi/2)$ phase difference between the $(E_x,\ E_y)$ field components are the C-point singularity. The polarization ellipse orientation and the phase difference are calculated and plotted from the measured Stokes parameters in Fig. 3(d) and 3(e). Figure 3(f) is the plot where the singularities are identified in the beam cross-section, where the beam intensity and the polarization ellipse orientations are plotted in the background. Red and green circles with a white dot correspond respectively to the C-point singularity in the right and left elliptical polarization fields, and the yellow line corresponds to the L-line singularity. Zooming in on positions around the C-points, one can see that the polarization variations form a star-like topological pattern in the corresponding polarization ellipse field. All these significant beam-field aspects, the azimuthal separation of $S_3$, the appearance of C-point and L-line singularities and star-like topological pattern in the beam cross-section, due to retroreflection of a nonparaxially focused linearly polarized Gaussian beam, is a manifestation of the underlying SOI of light. It is also important to note that the spatially separated right and left elliptical polarization on either side of the C-point (as can be seen in Fig. 3) corresponds to the photonic spin Hall effect (PSHE). The experimental results and the observed phase-polarization characteristics in the beam cross-section are confirmed via theoretical simulations, shown in Fig. 4 (a’, b’, c’, d’, e’, f’). As can be seen, while all the features mentioned above are also observed in the simulated results, deviations can be attributed to differences in the mirror characteristics used in the simulations.

Having established the topological features arising due to SOI in a retroreflected nonparaxially focused linearly polarized Gaussian beam, we now use a right-circularly polarized Gaussian beam. This is achieved by using a $45^o$ oriented quarter-wave plate (QWP) kept after the polarizer in the input beam (Fig. 2). The mirror is placed at the focal plane ($z = 0\ mm$) of the lens. The right and left circular polarized components of the output beam, the ($I_5,\ I_6$) Stokes images show Gaussian and Laguerre-Gaussian (LG) beams, as shown in Fig. 5 (a), (b). The LG beam has charge $l = +2$, due to spin-to-orbital AM conversion (SOC). The $S_3$ image calculated from these measurements, shown in Fig. 5 (c) demonstrates radial Hall effect of light [W. Shu, Y. Ke, Y. Liu, X. Ling, H. Luo, and X. Yin, Radial spin Hall effect of light, Phys. Rev. A 93, 013839 (2016)], the right and left elliptically polarized components of the beam are radially separated. The ellipticity variation in the beam cross-section and the phase difference between the transverse field components confirm the presence of a charge 2 vortex at the beam center (Fig. 5 (d), (e)). The small separation between the vortex centers allows the appearance of two C-point singularities in the beam center, surrounded by a circular L-line singularity. As can be seen, the polarization topology around the C-points is lemon/monstar-like (of charge $+1/2$) to maintain the conservation of AM. Merging the topological structures will produce a circular-polarized beam center with radial variation in beam intensity, as shown in the simulated results (Fig. 6 (a’, b’, c’, d’, e’, f’). In addition, all other simulated results of intensity, phase, and polarization variations in the beam cross-section also match the experimental results. Thus, using a circularly polarized Gaussian input beam, we demonstrate the corresponding SOI effects of spin-to-orbital AM conversion, radial PSHE, and C-point singularity with AM-conserving polarization topologies and an L-line singularity surrounding the C-points. All these intricate SOI features are enabled by retroreflecting a nonparaxially focused, polarized Gaussian beam off a high-reflectivity multilayer-coated dielectric mirror. While the measurements and simulations are carried out in the collimated output beam of ~0.8 mm size, back-projecting the output beam to the focal plane reduces the beam size to ~0.8 µm due to the 100X microscope objective lens. Further, the C-point singularity of ($5 \times 5\ \mu m$) pixel size, measured in the output beam cross-section, due to back projection, is actually of size 50 nm at the focal plane. We emphasize here that the retroreflection method demonstrated here enables experimental measurement of nano-scale phase-polarization variations in the beam cross-section.

By extending our ability to experimentally measure spatially resolved Stokes parameters and hence phase-polarization changes in the beam cross-section at the focal plane, we investigate the PSHE in 3D as a function of the mirror position in the focal region of the nonparaxial focus. For linear polarized Gaussian input beam, we translate the mirror from $-2\ \mu m$ to $+2\ \mu m$ in steps of $0.5\ \mu m$ and from the $(I_5,\ I_6)$ intensity measurements at each position, calculate and plot $S_3$ versus $z$ (Fig. 7). Each experimentally obtained $S_3$ image is normalized to maximum of total intensity $(S_0)$ value giving $s_3 = \frac{S_3}{S_{0max}}$. From the figure, we can see that before the focal plane, the SoP of the 4-lobe pattern is right elliptical in 1st and 3rd quadrants and left elliptical in the 2nd and 4th quadrants. The SoP of the 1st and 3rd quadrant lobes flips to right elliptical, and the 2nd and 4th quadrant lobes become left elliptical upon crossing the focal plane. The limitation in the minimum distance of travel by the mirror is compensated by filling the gap with the simulated results from $-0.4\ \mu m$ to $+0.4\ \mu m$ in steps of $0.1\ \mu m$. From Fig. 7 it is clear that the right and left elliptical SoP of the lobes interchange before and after focal plane along with the decrease and increase in the beam spot size as one approach the moves away from the focal plane. The experimental demonstration of PSHE along the propagation direction $(z -)$ confirms theoretical simulations of an earlier work [Ref]. Despite the flip in the polarization ellipticity measured via $S_3$, the AM variation across the focal plane remains unaltered, as can be seen from the total phase-polarization variation measured in the beam cross-section at three identified locations (Fig. 8 (a) – (c)). The polarization topologies continue to be star-like patterns across the focal plane, and the only difference between them is the spatial position where the singularities appear, satisfying the necessary condition.

The four lobes for a linearly polarized Gaussian beam are azimuthally spin separated as shown via the $S_3$ at all mirror distances. Besides, the presence of C-point singularities of the corresponding circular polarization in the respective lobes, with a surrounding lemon- or star-like pattern, leads to PSHE due to the spin-phase gradient. Thus, one would expect to see the effect of the presence of the spin-phase gradient between any pair of lobes, leading to a transverse spin shift of the lobes. However, unlike in situations where the effects are very small, as around the C-point singularities, which require measurements via the weak measurement protocol, the enhanced SOI effects allow us to perform standard polarimetry measurements. After identifying the exact cross-polarization angle of the QWP-P2, giving the 4-lobe pattern with Maltese dark cross pattern (Fig. 9 (b)), the analyzer is rotated by 10º from -85º to +95º. The obtained patterns are shown in Figs. 9 (a) and (c). As can be seen, if the diagonal (1 and 3) lobes meet at the center, the anti-diagonal (2 and 4) lobes are pushed away from the beam center (Fig. 9(b)), due to spin-phase gradient. This is the transverse spin shift, which arises and becomes measurable due to the enhanced SOI characteristics of a retroreflected, nonparaxially focused, linearly polarized Gaussian beam. The joining and pushing away of the lobes flip with each other for 95º angle of the analyzer, due to the right and left elliptical polarization of the diagonal and anti-diagonal lobes. This allows us to measure the shift in the centroid position of each lobe as a function of the analyzer angle. For this, we identified a horizontal-vertical pair of lobes, as indicated in Fig. 9(b). Fig. 10 (a) and (b) gives the experimentally measured horizontal and vertical centroid shift respectively for the horizontal and vertical lobes marked in Fig. 9 (b), with the QWP kept fixed at 90º and P2 rotated by $90^{\circ} \pm \varepsilon$ with $\varepsilon = 5^{\circ}$. As can be seen from Fig. 10 (a), for horizontal lobes, the centroid shift along $y-$ direction increases linearly crossing zero shift for $\theta_A = 0^{\circ}$, and along the $(x-)$ direction the centroid shift is maximum at $\theta_A = \pm 5^{\circ}$ and zero shift for $\theta_A = 0^{\circ}$. The centroid shift of the lobes with analyzer angle reverses their behaviour for the vertical lobes (Fig. 10 (b)). Moving the mirror across the focal plane to positions before and after it, we measure the centroid shift for the horizontal lobes and show that while the $x-$ shift changes its slope (from negative to positive) across the focal plane, the $y-$ shift behavior remains the same as at the focal plane shown in Fig.10 (c) and 10 (d) . This is another confirmation that the SOP of the lobes flips from right-to-left elliptical as the beam is sampled by moving the mirror across the focal plane. As mentioned before, small variations between the theoretical and experimental behaviour are attributed to the lack of layer information for the mirror used and to intensity non-uniformity both between and within the mode structures used to measure the spin shift of the modes.

**Conclusion:**

We demonstrate a method to significantly enhance the SOI effects and the weak phase-polarization modifications in the cross-section of a nonparaxially focused Gaussian beam by using a dielectric multilayer-coated mirror. The angle-of-incidence and polarization-dependent mirror reflectivity modify the retroreflected beam characteristics, enabling us to uniquely identify polarization singularities in the focal region. These sub-wavelength features are used to track the phase-polarization modifications across the focal region. Apart from this, we also report experimental measurement of the spin-Hall effect of light across the focal region, known as the 3D-SHEL. Modeling the retroreflected beam characteristics using the vectorial angular spectrum method, combined with the mirror's complex Fresnel reflection coefficients, allows us to understand and explain its overall behavior. The superposition of enhanced optical near-fields and the lossless, broadband nature of the mirror reflectivity are expected to expand the technique's capabilities to modify and modulate the phase-polarization characteristics of the reflected beam, thereby realizing complex structures within the beam, including the generation of Skyrmions, Modius strips, twists, and knots.

**Figure caption:**

**Figure 1:** Schematic of the experimental setup. P1: GT polarizer, H/QWP: Half/Quarter-wave plate, BS: Beam splitter, MO: Micro objective lens, DM: Dielectric mirror, P2: Polarizer, Z0: defocus length, CCD: Camera.

**Figure 2:** Experimentally measured (a) – (c), and simulated (a') – (c') S3 plots as the mirror is scanned across the focal plane for (a) z = -1 μm, (b) z = 0 μm, and (c) z = 1 μm. Red rectangular boxes indicate the pixelated area used to compute the spatially averaged S3 value shown in (d). (d) S3 vs. z-plot (black line curve is simulated data, and black hollow circles are experimentally measured data).

**Figure 3:** (a) and (b) Right circular (RC) and left circular (LC) component measured by keeping QWP 90 and analyser 45 & 135 respectively. (c) Normalized s3 plotted from measured from RC and LC components. (d) ellipse orientation (e) phase plot of s12 (arg(S1 + i S2)) plotted from the measured stokes parameters (f) stream line plot with total intensity plot and polarization ellipse plot in the background with the c-points (Red colour circles with white dot correspond to RC and Green colour circles with white dot correspond to LC c-points and yellow circle shows l-lines)).

**Figure 4:** Simulated (a) and (b) Right circular (RC) and left circular (LC) components. (c) Normalized s3 plotted from calculated from RC and LC components. (d) ellipse orientation (e) phase plot of s12 (arg(S1 + i S2)) plotted from the calculated stokes parameters (f) stream line plot with total intensity plot and polarization ellipse plot in the background with the c-points (Red colour with white dot correspond to RC and Green colour with white dot correspond to LC c-points and yellow concentric circle shows l-lines).

**Figure 5:** (a) and (b) Right circular (RC) and left circular (LC) component measured by keeping QWP 90 and analyser 45 & 135 respectively. (c) Normalized s3 plotted from measured from RC and LC components. (d) ellipse orientation (e) phase plot of s12 (arg(S1 + i S2)) plotted from the measured stokes parameters (f) stream line plot with total intensity plot and polarization ellipse plot in the background with the c-points (Red colour circles with white dot correspond to RC c-points and yellow circle shows l-line)).

**Figure 6:** Simulated (a) and (b) Right circular (RC) and left circular (LC) components. (c) Normalized s3 plotted from calculated from RC and LC components. (d) ellipse orientation (e) phase plot of S12 (arg(S1 + i S2)) plotted from the calculated stokes parameters (f) stream line plot with total intensity plot and polarization ellipse plot in the background with the c-points (Red colour with white dot correspond to RC c-points and yellow circle shows l-line).

**Figure 7:** Experimentally measured (upper row) and simulated (lower row) S3 plots across the focal plane.

**Figure 8:** Stream lines plot with ellipse orientation plot and polarization ellipse plot in the background when mirror was scanned across the focus with step size 0.5 micro m (a) Before focus (z = -0.5 micro m) (b) At focus (z = 0micro m) (c) After focus (z = 0.5 micro m). (Here, white circles with yellow dot shows LC c-points and black circles with yellow dot shows RC c-points).

**Figure 9:** Weak measurement for the post selected orthogonal component (i.e. x-component) for the analyser angle (a) 85° (b) 90° (c) 95° (d) Animated video when the analyzer angle was varied from 85° - 95° with the step size 0.1° (white dot in each lobe shows center of mass of respective lobe).

**Figure 10:** Simulated line curves and experimentally measured hollow circles (a) X-shift (black colour) and Y-shift (Red colour) (b) Y-shift (black colour) and X-shift (Red colour) when mirror was kept at the focus for the horizontal and vertical two lobes respectively (see Fig. 8(b)). (c) X-shift and (d) Y-shift, when mirror was scanned across the focus, before focus (Red colour) , At focus (black colour) and after focus (Green colour) for horizontal two lobes.

**Figure 1**

CCD
P2
QWP
P1
LASER
BS
H/QWP
MO
(-)
$Z_0$
DM
(+)

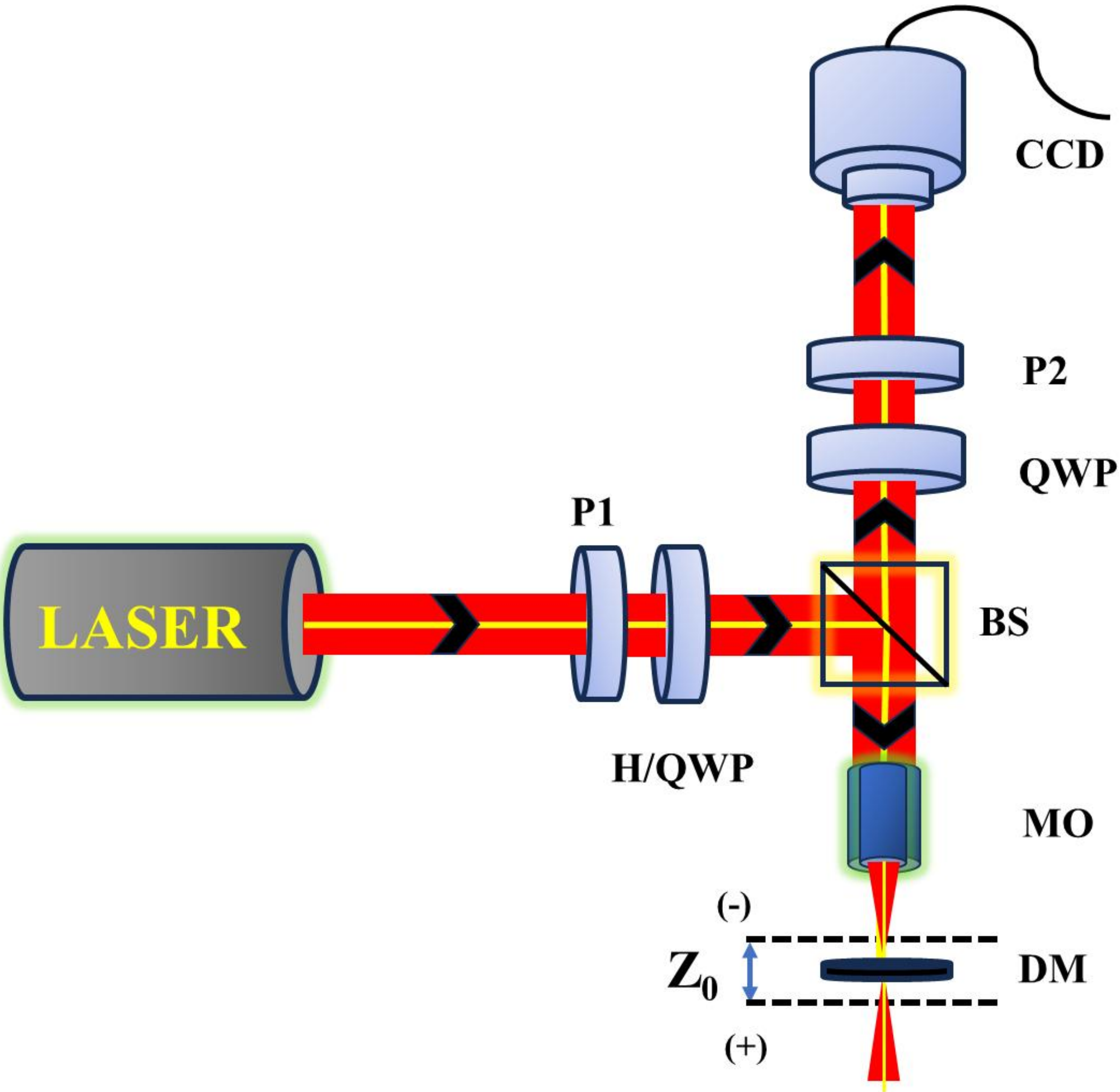

**Figure 2**

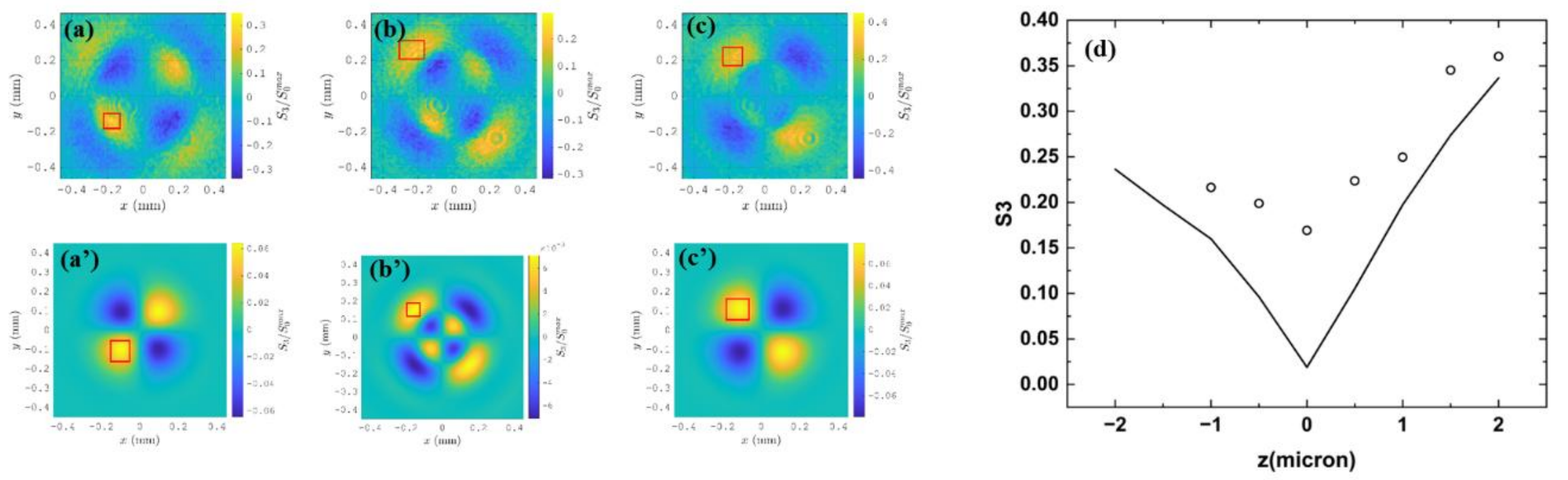

## Figure 3

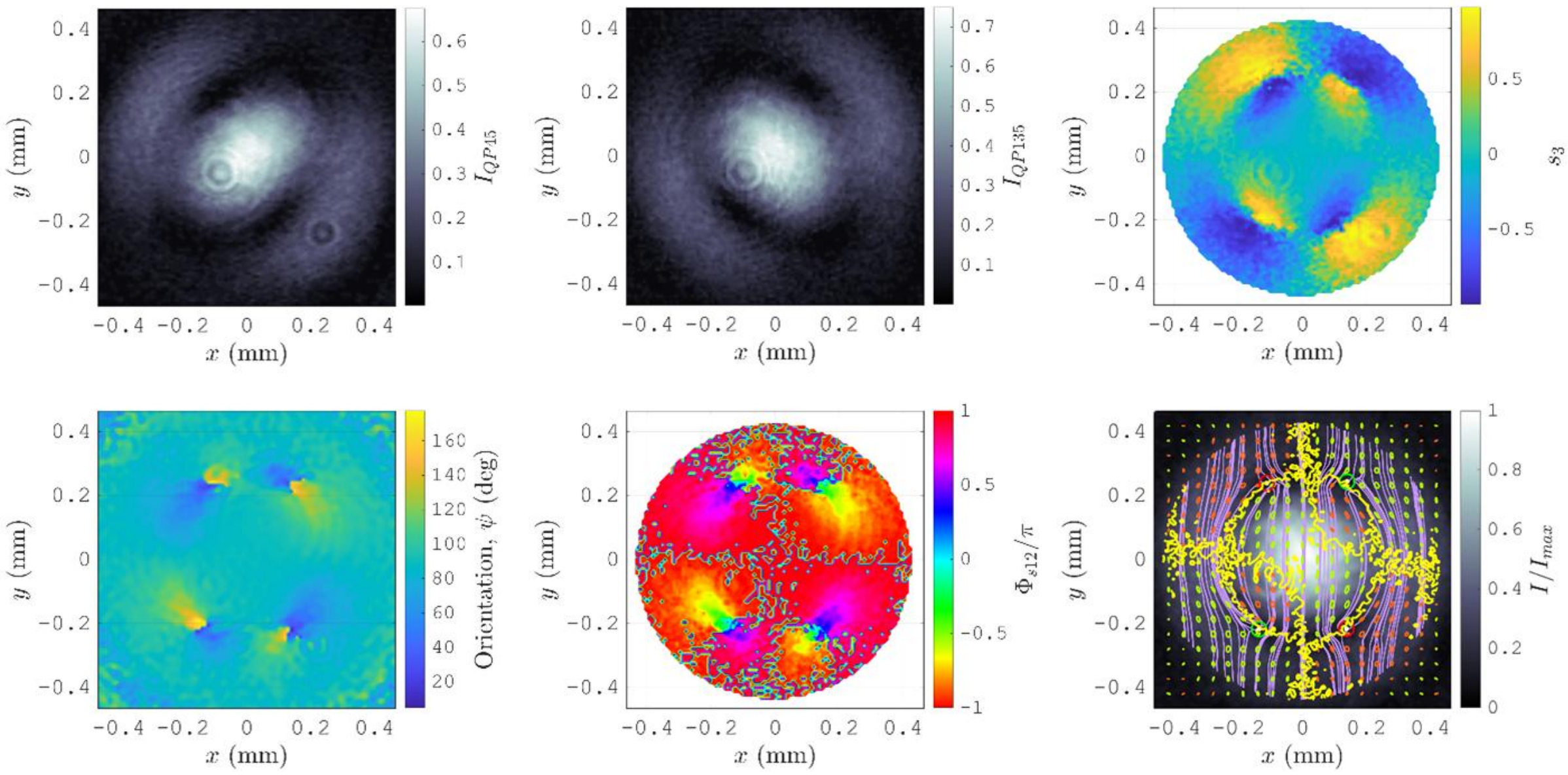

## Figure 4

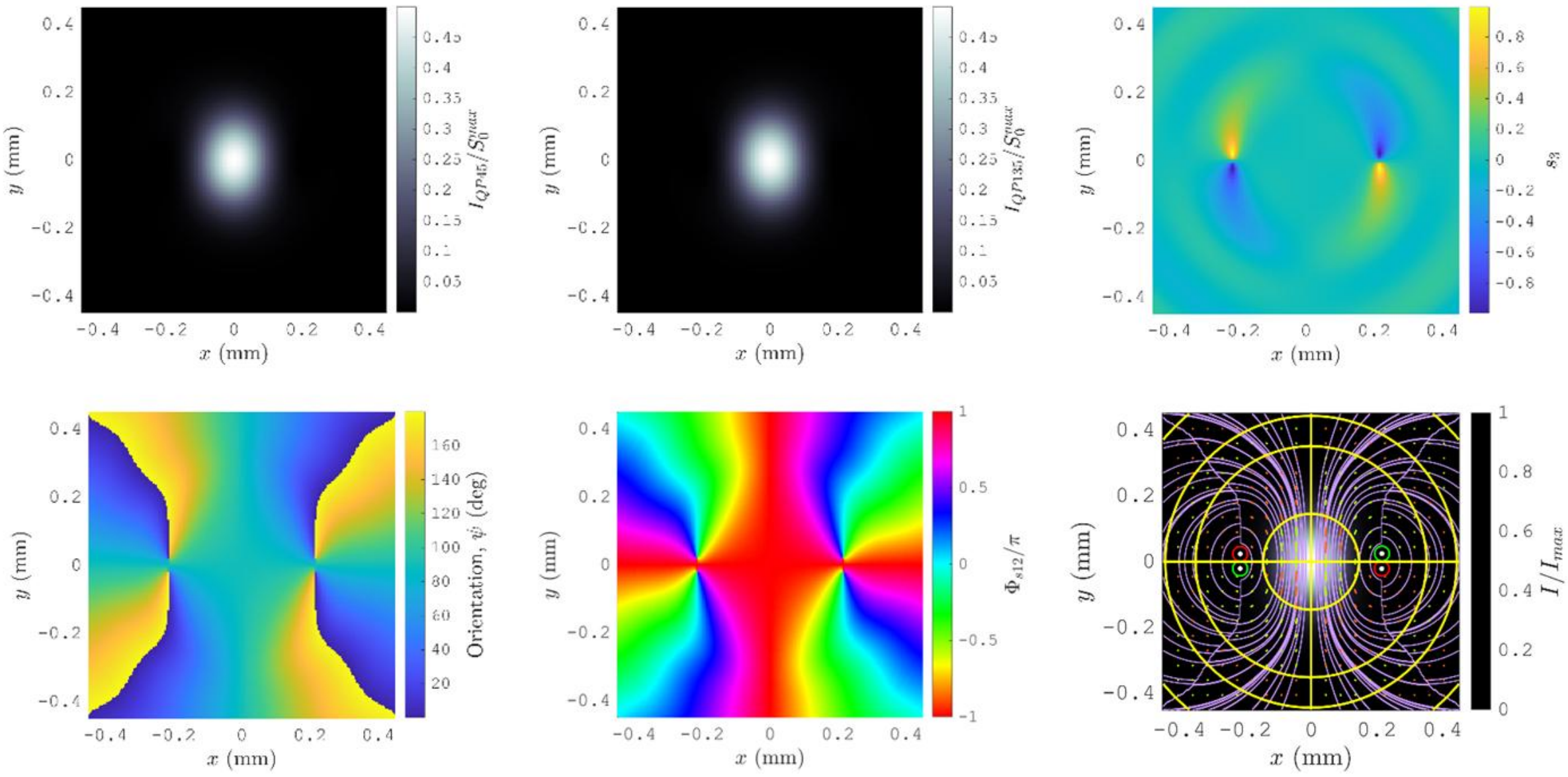

## Figure 5

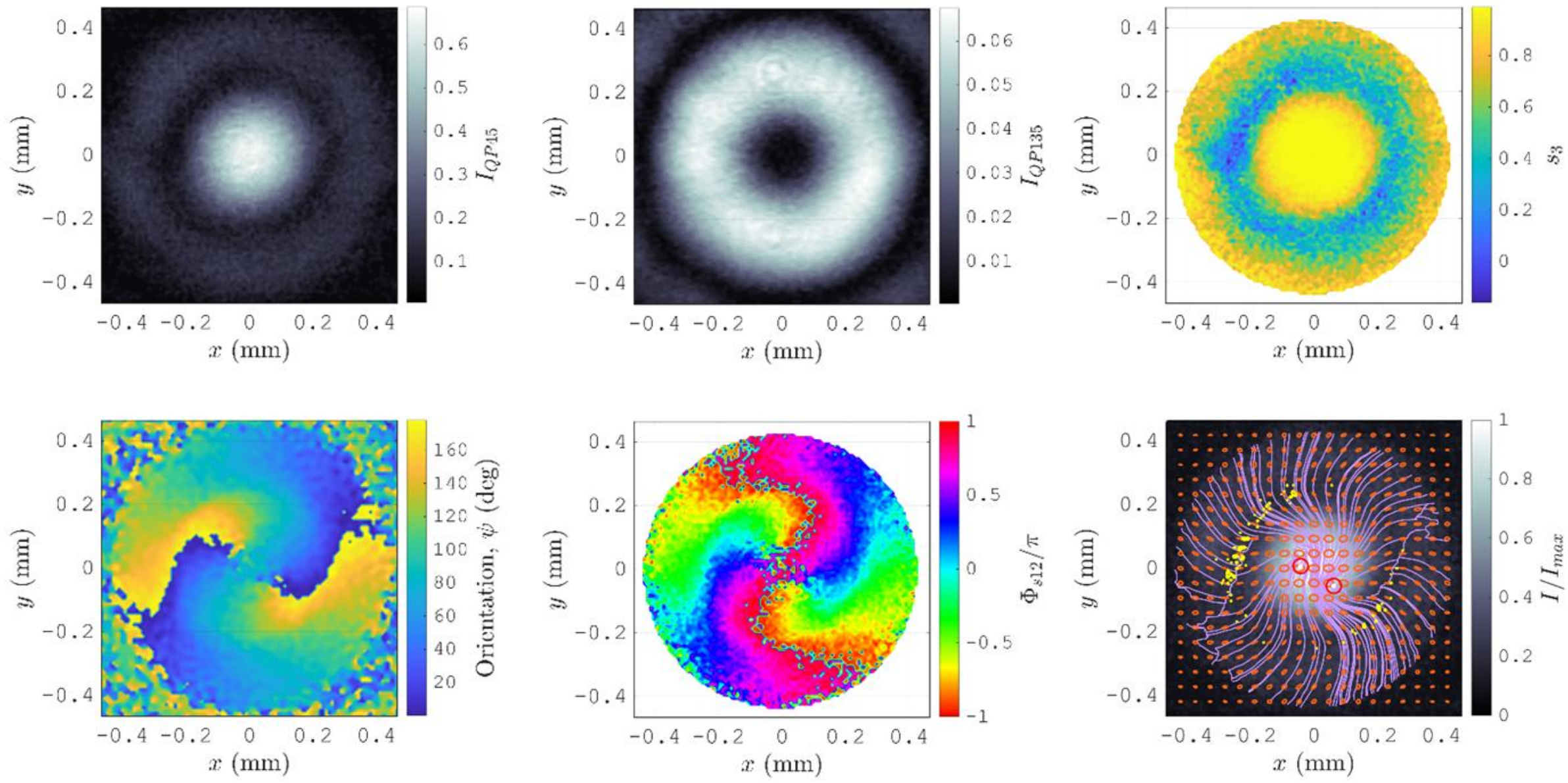

## Figure 6

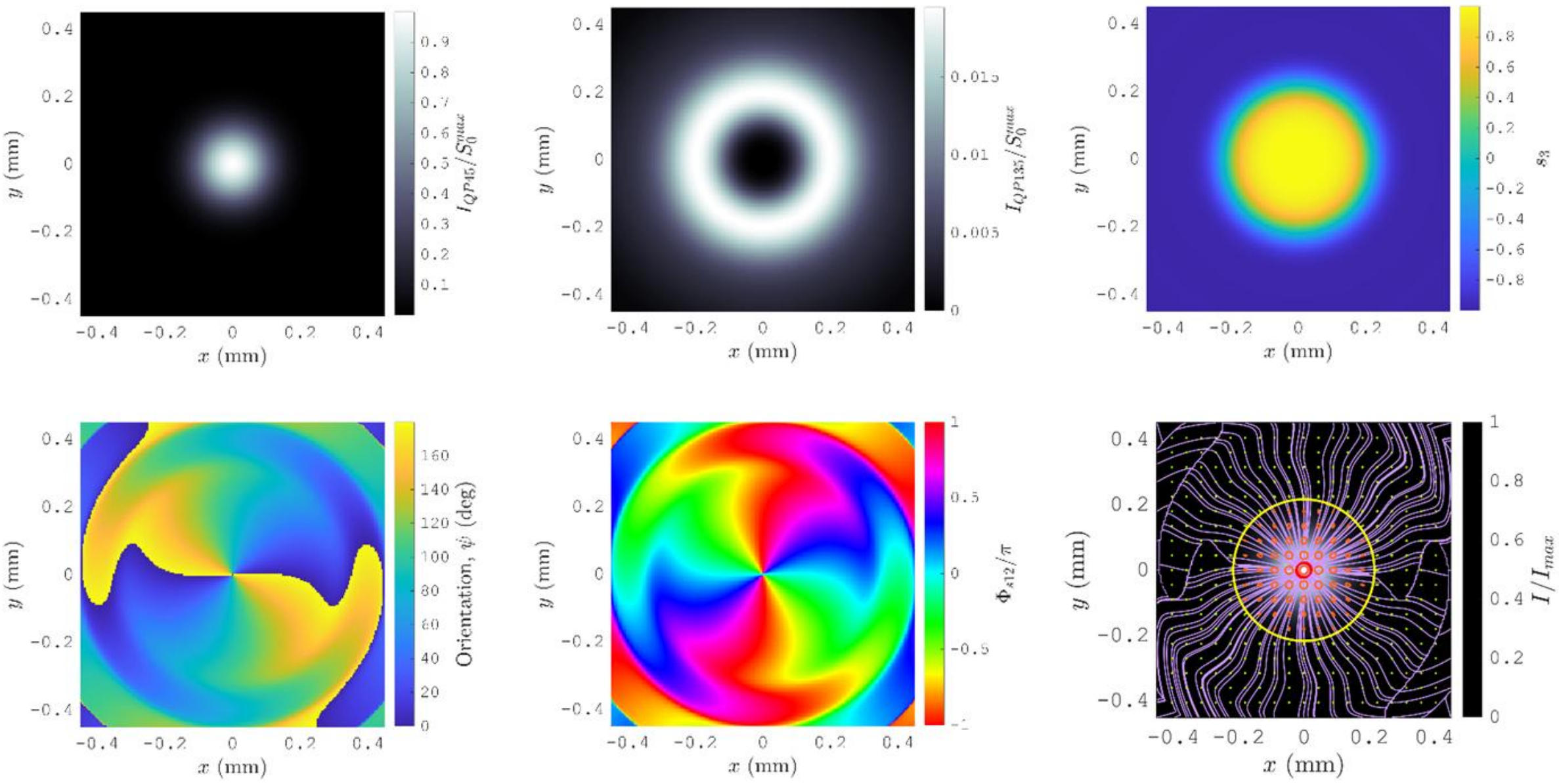

**Figure 7**

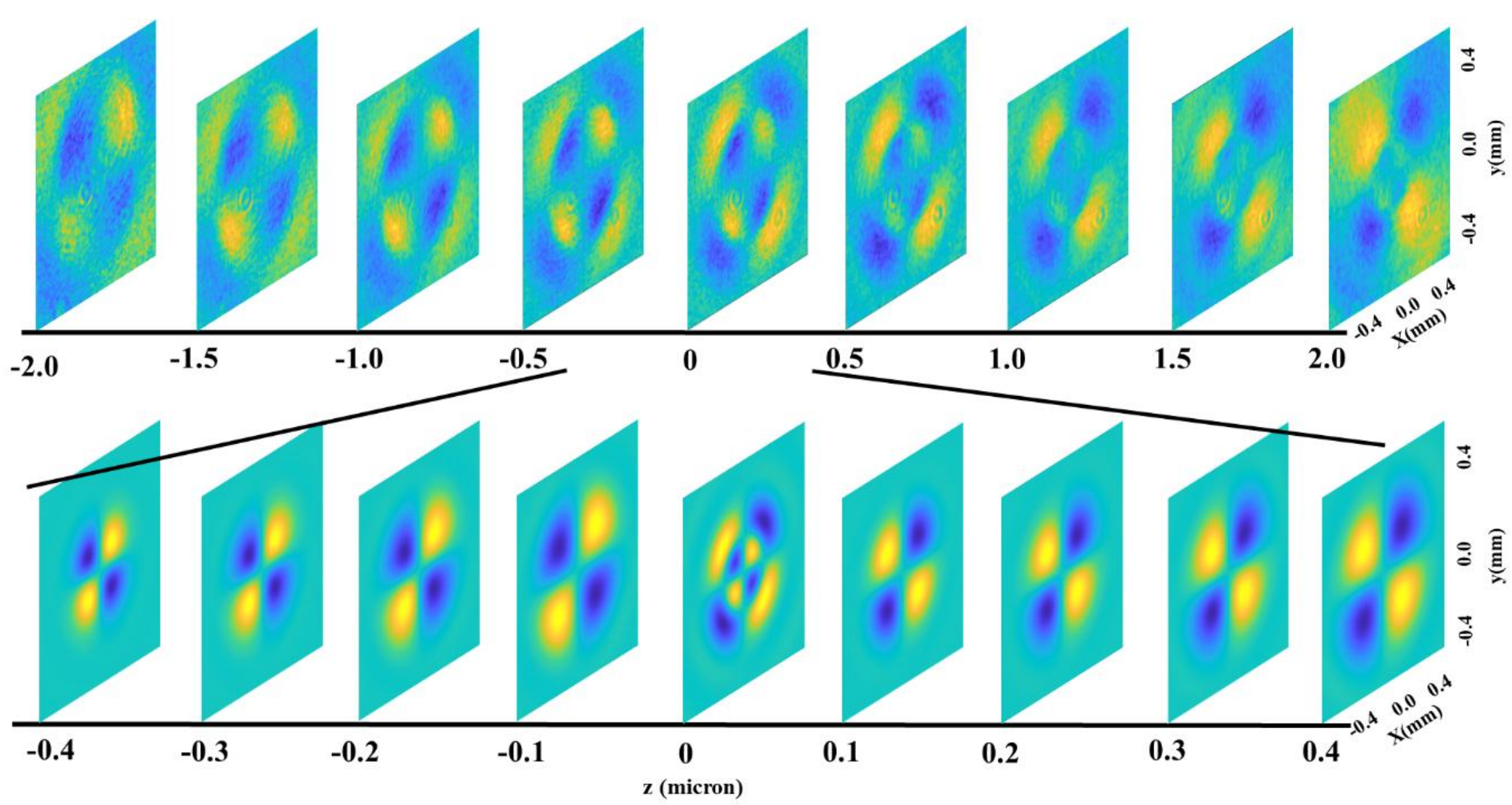

**Figure 8**

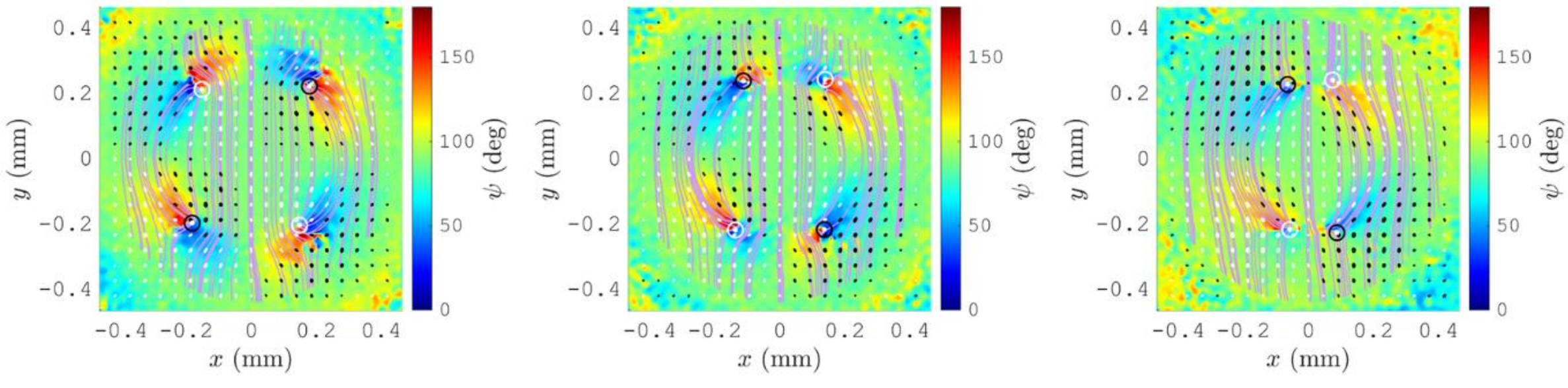

**Figure 9**

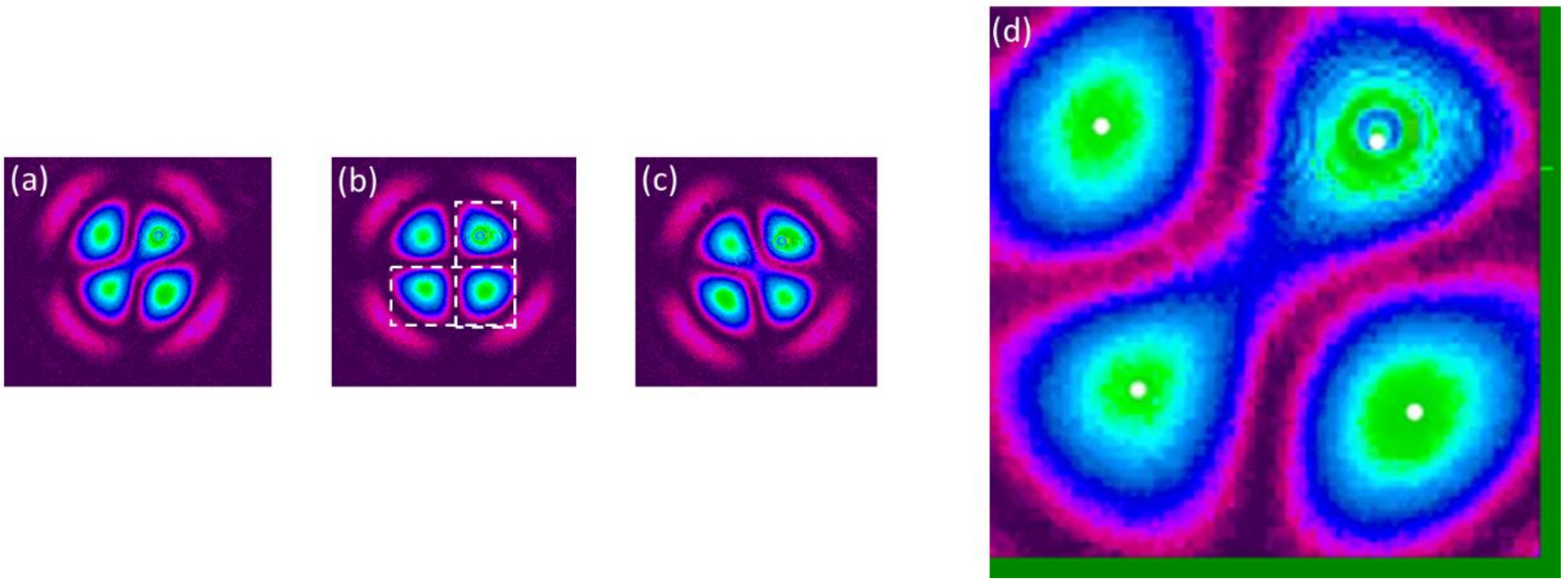

**Figure 10**

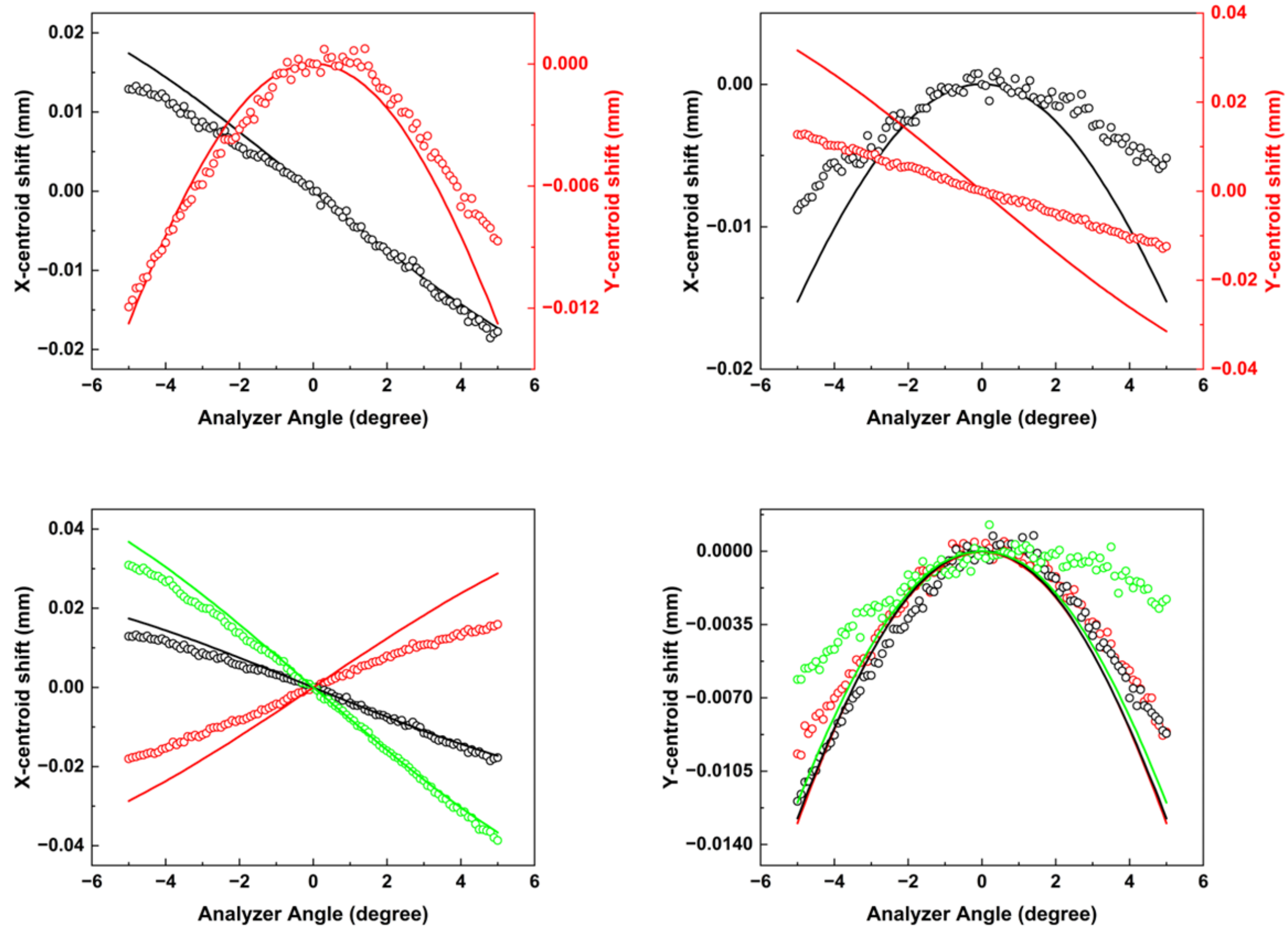

X-centroid shift (mm)
Y-centroid shift (mm)
Analyzer Angle (degree)
0.02
0.01
0.00
−0.01
−0.02
0.000
−0.006
−0.012
−0.6
−4
−2
0
2
4
6
0.04
−0.04
−0.0140
−0.0105
−0.0070
−0.0035
0.0000